\documentclass[aps,preprint,showpacs]{revtex4}
\usepackage{float}
\usepackage{amsmath}
\usepackage{amssymb}

\begin{document}

\title{Stable three-dimensional solitons in attractive Bose-Einstein
condensates loaded in an optical lattice}
\author{D. Mihalache$^{1,2,3}$, D. Mazilu$^{2,3}$, F. Lederer$^{2}$, B. A.
Malomed$^{4}$, L.-C. Crasovan$^{1,3}$, Y. V. Kartashov$^{1}$, and
L. Torner$^{1}$} \affiliation{$^{1}$ICFO-Institut de Ciencies
Fotoniques, and Department of Signal Theory and Communications,
Universitat Politecnica de Catalunya, 08034 Barcelona, Spain}
\affiliation{$^{2}$ Institute of Solid State Theory and
Theoretical Optics, Friedrich-Schiller Universit{\"a}t Jena,
Max-Wien-Platz 1, D-077743 Jena, Germany}
\affiliation{$^{3}$National Institute of Physics and Nuclear
Engineering, Department of
Theoretical Physics, P.O. Box MG-6, Bucharest, Romania}
\affiliation{$^4$Department of Interdisciplinary Studies, Faculty
of Engineering, Tel Aviv University, Tel Aviv 69978, Israel}

\begin{abstract}
The existence and stability of solitons in Bose-Einstein condensates with
attractive inter-atomic interactions, described by the Gross-Pitaevskii
equation with a three-dimensional (3D) periodic potential, are investigated
in a systematic form. We find a one-parameter family of stable 3D solitons
in a certain interval of values of their norm, provided that the strength of
the potential exceeds a threshold value. The minimum number of $^{7}$Li
atoms in the stable solitons is $60$, and the energy of the soliton at the
stability threshold is $\approx 6$ recoil energies in the lattice. The
respective energy-vs.-norm diagram features two cuspidal points, resulting
in a typical \textit{swallowtail pattern}, which is a generic feature of 3D
solitons supported by low- (2D) or fully-dimensional lattice potentials.
\end{abstract}

\maketitle


Creation of multidimensional solitons built of nonlinear light or matter
waves is a great challenge to the experiment. The current situation in this
field is summarized in a recent review \cite{JOptReview}. The only real
example of a quasi-two-dimensional (quasi-2D) spatiotemporal soliton in
optics was one created in a $\chi ^{(2)}$ (quadratically nonlinear) crystal
\cite{Frank}; it could not be made fully three-dimensional, as one
transverse direction was reserved to implement the technique of tilted wave
fronts, by means of which sufficiently strong artificial dispersion was
induced in the medium.

While the experimental situation in optics remains difficult
\cite{JOptReview}, new possibilities for the creation of
multidimensional solitons are offered by Bose-Einstein condensates
(BECs) with attractive interactions between atoms. In an
effectively 1D condensate of $^{7}$Li, single solitons and soliton
clusters were successfully created \cite{Li}. In the 2D and 3D
situations, the self-focusing cubic nonlinearity induced by the
attractive interactions leads, respectively, to weak and strong
collapse, as predicted by the corresponding Gross-Pitaevskii
equation (GPE) [alias the nonlinear Schr\"{o}dinger (NLS)
equation] \cite{Berge}, and demonstrated experimentally in the BEC
\cite{collapse-experiment}. However, a spatially periodic
potential in the form of an optical lattice (OL), created as an
interference pattern by coherent laser beams illuminating the
condensate, can stabilize the multidimensional solitons in the
self-attractive BEC. For the 2D case, this was first predicted in
Refs. \cite{BBB1,Yang}; moreover, it was also demonstrated
\cite{BBB2} that 2D solitons may be stable in the presence of a
quasi-1D periodic potential (one that does not depend on the
second spatial coordinate). 3D solitons supported by the 3D OL
were reported too \cite{BBB1}. The form of the soliton was
predicted by the variational approximation (VA), which was used,
as an initial guess, to generate several examples of stable 3D
solitons in direct simulations, including ``single-cell" and
``multi-cell" ones, that are essentially confined to one or
several cells of the OL, respectively. Further, in Refs.
\cite{BBB2} and \cite{MihalachePRE2004} it was independently
demonstrated that stable 3D solitons can also be supported by the
low-dimensional, i.e., quasi-2D, lattice potential (however, the
quasi-1D potential cannot stabilize 3D solitons \cite{BBB2},
unless it is combined with periodic alternation of the
nonlinearity sign, provided by the Feshbach resonance in ac
magnetic field \cite{Warsaw}). In particular, the VA predicts that
the 3D solitons in the quasi-2D lattice exist at all values of the
norm (number of atoms) $N$ and OL strength $p$, but are stable
only for $p$ exceeding a certain critical value $p_{\mathrm{cr}}$,
in an interval of the width $\Delta
N_{\mathrm{stab}}^{\mathrm{(3D)}}\sim (p-p_{\mathrm{cr}})^{1/2}$ in
a vicinity of a finite value $N_{\mathrm{cr}}$ of the norm
\cite{BBB2} [the stability was predicted on the basis of the
Vakhitov-Kolokolov (VK) \cite{VK} criterion].

Except for a few examples reported in Ref. \cite{BBB1}, no systematic
investigation of the existence, stability and robustness of 3D BEC solitons
in the 3D lattice potential has been performed yet. The objective of the
present work is to report basic results for this problem.

The GPE provides for the description of the BEC dynamics in terms of the
mean-field single-atom wave function $\psi (x,y,z,t)$ \cite{GP}. The
normalized form of this equation for a self-attractive condensate trapped in
the 3D potential $-V(x,y,z)$ is well known \cite{BECreview}:

\begin{equation}
i\frac{\partial \psi }{\partial t}=-\frac{1}{2}\left(
\frac{\partial ^{2}\psi }{\partial {x}^{2}}+\frac{\partial
^{2}\psi }{\partial {y}^{2}}+\frac{\partial ^{2}\psi }{\partial
{z}^{2}}\right) -|\psi |^{2}\psi -V(x,y,z)\psi   \label{evolution}
\end{equation}Generally, $V(x,y,z)$ contains terms accounting for the confining parabolic
trap (magnetic and/or optical) and the periodic potential of the
OL. Being interested in the localized solutions, occupying a few
cells of the lattice, we disregard the parabolic potential and set
$V(x,y,z)=p[\mathrm{cos}(4x)+\mathrm{cos}(4y)+\mathrm{cos}(4z)]$,
where the OL period is normalized to be $\pi /2$, and the OL
strength $p$ is defined to be positive. Besides the
above-mentioned norm, $N~=~\int \int \int \left\vert \psi
(x,y,z)\right\vert ^{2}dxdydz$, Eq. (\ref{evolution}) conserves
the energy E (see \cite{BECreview}).
Stationary soliton solutions have the form $\psi
(x,y,z,t)=w(x,y,z)\exp (-i\mu t)$, with a real function $w$ and
chemical potential $\mu $. We looked for the function $w(x,y,z)$
by means of the known method of the propagation in imaginary time
\cite{IT}. It was implemented, using the Crank-Nicholson scheme,
with the nonlinear finite-difference equations solved by means of
the Picard iteration method, and the resulting linear system
handled with the help of the Gauss-Seidel iterative procedure. To
achieve good convergence, we typically needed six Picard
iterations and six Gauss-Seidel iterations. We used equal
transverse grid stepsizes, $\Delta x=\Delta y=\Delta z\equiv h$,
and a mesh of $361\times 361\times 361$ points was usually
employed. The convergence to a stationary state occurred after
$4\times 10^{3}-5\times 10^{4}$ steps of the evolution in
imaginary time, typical transverse-grid and time stepsizes being
$h=0.02$ and $\Delta t=0.0003$, respectively, for narrow solitons,
whereas for broad ones it was enough to take $h=0.07$ and $\Delta
t=0.004$.

One can derive a relationship between the norm $N$, chemical
potential $\mu $, real wave function $w$ and energy $E$ of the
stationary solution: $E=\mu N+\frac{1}{2}\int \int \int
w^{4}(x,y,z)dxdydz$. It can be used to determine the chemical
potential $\mu $, once the field profile $w$ is known. This exact
relation was also used for verification of accuracy of the
numerically found stationary solutions. Notice that, for
stationary solitons of the NLS equation in the free 3D space [with
$V=0$ in Eq. (\ref{evolution})], the following relations between
$\mu $, $N$, and $E$ are known: $\mu
(N)=-CN^{-2},\,E(N)=CN^{-1},\,C\approx 44.5$ \cite{Yuri,AA} (a
corollary of this is $d\mu /dN>0$, which immediately shows that
these free-space solitons are \emph{always unstable}, as the VK
stability criterion requires exactly the opposite, $d\mu /dN<0$).

In Figs. 1(a) and 1(b) we plot the dependences $\mu =\mu (N)$ and $E=E(N)$
for the numerically found family of 3D solitons in the present model. It is
seen that, in the presence of the 3D OL, the localized states exist only for
$\mu $ smaller than some maximum value, $\mu _{\max }(p)$; in fact, it
corresponds to the edge of the \textit{bandgap} in the spectrum of the
linearized equation (\ref{evolution}). At values of $\mu $ that do not
belong to the bandgap, no soliton is possible \cite{Yuri}. We note that $\mu
_{\max }(p)$ decreases with the increase of the lattice strength $p$, see
Fig 1(a) [for the 3D NLS equation in free space ($p=0$), one has $\mu _{\max
}(0)=0$].
Remarkably, for sufficiently large values of the lattice strength
$p$, the $E(N)$ curves in Fig. 1(b) feature \textit{two cusps},
instead of a single one, as in most other 2D and 3D Hamiltonian
models. Examples of the usefulness of the energy-vs.-norm diagrams
(or Hamiltonian-vs.-power diagrams, in the context of spatial
optical solitons) in the analysis of the existence and stability
of solitons can be found in Ref. \cite{AA}.
To check the soliton's stability 
by direct propagation simulations
we used the standard Crank-Nicholson
discretization scheme with $121\times 121\times 121$ points in the
coordinates $x,y,z$, and spatial and time stepsizes $0.019\leq h\leq 0.073$
and $0.00045\leq \Delta t\leq 0.004$ (depending on the soliton's size).
The validity of the VK stability criterium was checked by performing direct 
propagation simulations;
the step in the soliton's parameter was $\Delta \mu=0.08$.
Thus, we
have found stable 3D solitons, in a finite interval of the values of $N$, if
the OL strength $p$ exceeds a threshold value $p_{\mathrm{cr}}$: as seen
from Fig. 1, $p_{\mathrm{cr}}$ itself is located between $p=1.5$ and $p=2$.
Further, Fig. 2 displays an integrated characteristic of the
soliton family -- the dependence $E=E(\mu ,N)$ -- together with
the corresponding projections onto the three planes $(E,\mu )$,
$(E,N)$, and $(\mu ,N)$, for two different values of the lattice
strength, $p=0$ and $p=3$. In the latter case [in Fig. 2(b)], we
notice the \textit{non-monotonic} behavior of the three projected
curves for $p=3$ (above the stability threshold) and the
\textit{``swallowtail" loop} in the energy-vs.-norm diagram.
Although this pattern is one of generic possibilities known in the
catastrophe theory, it rarely occurs in physical models
(applications of the catastrophe theory to the soliton-stability
problem were reviewed in Ref. \cite{Kusmartsev}).

It is noteworthy that the qualitative features of the stability picture for
the 3D solitons are essentially the same as found earlier in the approximate
analytical form by means of the VA \cite{BBB2}, and in the numerical form as
well \cite{BBB2,MihalachePRE2004}, for 3D solitons supported by the
low-dimensional (quasi-2D) lattice: a narrow stability interval $\Delta N$
appears in a vicinity of a finite critical value $N_{\mathrm{cr}}$, when the
lattice strength $p$ exceeds a finite minimum value $p_{\mathrm{cr}}$.
Remarkably, the \textit{``swallowtail" loop} was also identified
as a characteristic feature of the family of stable 3D solitons
supported by 2D harmonic lattices \cite{MihalachePRE2004}.
It is relevant to compare the stability picture for the 3D
solitons supported by a fully-dimensional OL with that for 2D
solitons supported by an OL (that may be either a low-dimensional
quasi-1D lattice, or the full 2D one) \cite{BBB1,BBB2,Moti}: in
that case stable solitons appear at \emph{arbitrarily small}
values of the lattice strength, and their stability interval
extends, in terms of $N$, up to a maximum (``cutoff") value at
which the 2D solitons exist (the latter is actually the norm of
the \textit{Townes soliton} i.e., a soliton of the free-space 2D
NLS equation \cite{Berge}). Thus, we conclude that all the above
features, \textit{viz}., the existence of stability threshold
$p_{\mathrm{cr}}$, the (small) stability interval $\Delta
N_{\mathrm{stab}}$ for $p>p_{\mathrm{cr}}$, and the typical
swallowtail pattern in the dependence $E=E(N)$, are generic to 3D
solitons supported by lattice potentials, and distinguish them
from 2D solitons.

Shapes of both unstable and stable solitons are shown in Fig. 3,
through their isosurface plots for a typical value of the lattice
strength parameter, $p=3$. An unstable low-amplitude 3D soliton
(with $A\equiv w(0,0,0)=1.8$), found for the value of the chemical
potential $\mu $ close to the bandgap edge, is displayed in Fig.
3(a). This soliton is broad, occupying many lattice cells. Typical
\emph{stable} solitons, with medium ($A=2.2)$ and high ($A=3$)
amplitude, are shown in Figs. 3(b) and Fig. 3(c), respectively.
Notice that the unstable soliton in Fig. 3(a) and its stable
counterpart in Fig. 3(c) have \emph{equal values of the norm},
$N=2.4$. The intermediate stable soliton in Fig. 3(b) has a
smaller norm ($N=2.04$), which is very close to the limit value
corresponding to the first cuspidal point in the dependence
$E=E(N)$, see the curve pertaining to $p=3$ in Fig. 1(b). Figure 4
additionally shows integrated views along the $z$-axis of the
isosurface plots displayed in Fig. 3. This figure illustrates the
fact that low-amplitude solitons spread to many lattice cells,
whereas high-amplitude solitons occupy only a few cells.

An important issue for these 3D solitons is occurrence of the
collapse (recall that the 3D collapse in the NLS/GPE is strong, on
the contrary to the weak 2D collapse \cite{Berge}). We expect that
solitons on the stable branch are able
to withstand small perturbations without collapse, whereas
linearly unstable solitons would either collapse, or reshape
themselves into time-periodic breathers, or decay into
``radiation" (quasi-linear Bloch waves), depending on the type and
strength of the perturbation. In order to verify these
expectations, we simulated the evolution of the solitons under
small
perturbations, taking the initial condition as 
$\psi (t=0)=w(x,y,z)(1+\epsilon \rho )$, 
where $\epsilon $ is a small amplitude of the perturbation, and
$\rho $ is taken either as a random number from the interval
$[-0.5,0.5]$ (stochastic perturbation) or as $\rho \equiv 1$
(uniform perturbation). We have checked that the solitons 
belonging to the VK-stable segments of
the curve $\mu =\mu (N)$ are indeed stable against small perturbations.
To illustrate this result, Fig. 5 shows an example of a stable
soliton which persists after the application of the stochastic
perturbation with $\epsilon =0.1$: in the course of the evolution,
the soliton's amplitude slightly oscillates, with no trend to
collapse or breakup. This and many other simulations clearly show
that the linearly stable solitons are also \textit{nonlinearly
stable} objects.
For linearly unstable solitons, the simulations reveal the
following scenarios of the instability development. (i)
Low-amplitude solitons decay into linear waves under stochastic
perturbations, or under uniform ones that reduce the soliton's
norm, i.e., perturbations with $\rho \equiv 1$ and $\epsilon <0$;
this case is illustrated by Fig. 6 for a typical situation. (ii)
The same unstable soliton, but under uniform perturbations with
$\rho \equiv 1$ and $\epsilon >0$ (that make its norm larger),
reshapes itself into a time-periodic breather. (iii) An unstable
high-amplitude soliton collapses if it is perturbed by the uniform
perturbation that increases its norm.

In conclusion, we have found one-parameter soliton families of the 3D
nonlinear Schr\"{o}dinger/Gross-Pitaevskii equation with self-focusing
nonlinearity and 3D lattice potential, and explored their stability.
Comparing the results with recently published findings for stable 3D
solitons in the quasi-2D lattice potential, and with (very different)
results for 2D solitons supported by the quasi-1D or full 2D lattice, we
were able to identify generic features that distinguish the stable 3D
solitons: a finite stability threshold $p_{\mathrm{cr}}$ in terms of the
lattice strength $p$, a finite stability interval $\Delta N$ in terms of the
soliton's norm $N$, and the swallowtail shape of the energy-vs.-norm
dependence.
The 3D solitons investigated here can be created in BEC -- most
plausibly, using $^{7}$Li loaded into a 3D optical lattice, with
the spatial period $\Lambda \sim 0.5~\mu $m. Then, undoing
normalizations that cast the GPE into the rescaled form
(\ref{evolution}), it is easy to find that the actual number of
atoms $N_{\mathrm{phys}}$ is related to the solution norm by
$N_{\mathrm{phys}}=\left( 2\pi ^{2}|a|\right) ^{-1}\Lambda N$,
where $a$ is the scattering length of the atomic collisions. For
$^{7}$Li, $a=-1.45$ nm, which yields the minimum number of atoms
necessary for the formation of stable solitons, that corresponds
to $N_{\mathrm{cr}}\approx 3.5$ in Fig. 1:
$N_{\mathrm{phys}}^{(\min )}\approx \allowbreak 60$ (note that it
corresponds to the density $\sim 5\times 10^{14}$ cm$^{-3}$, which
is quite a typical value for BEC). The strength $\varepsilon $ of
the OL, chemical potential, and energy in physical units are
related to their normalized strength counterparts as
$
\varepsilon =(1/8)E_{\mathrm{rec}}p,~\mu
_{\mathrm{phys}}=(1/4)E_{\mathrm{rec}}\mu
,~E_{\mathrm{phys}}=[\hbar ^{2}/(8|a|\Lambda m)]E\equiv
\{\Lambda /[(4\pi )^{2}|a|]\}E_{\mathrm{rec}}E$,  
where $E_{\mathrm{rec}}=(2/m)(\pi \hbar /\Lambda )^{2}\approx \allowbreak
7.\,5\times 10^{-29}$ J is the recoil energy in the lattice. Thus,
near the soliton-stability threshold, 
we get 
$\varepsilon \approx 0.25E_{\mathrm{rec}}$, $\mu
_{\mathrm{phys}}\approx -E_{\mathrm{rec}}$, and $E_{\mathrm{phys}}\approx
\allowbreak -6E_{\mathrm{rec}}$.

Support from Instituci{\'{o}} Catalana de Recerca i Estudis Avan{\c{c}}ats
(ICREA), Barcelona, Deutsche Forschungsgemeinschaft (DFG), and Israel
Science Foundation (grant No. 8006/03) is gratefully acknowledged.


\newpage



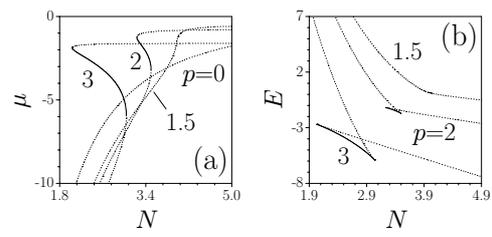
\begin{figure}[tbp]
\caption{The chemical potential $\protect\mu $ (a) and the energy $E$ (b) of
the 3D solitons versus their norm $N$ for different values of the lattice
strength $p$. Full and dotted lines depict stable and unstable solitons,
respectively.}
\label{Fig1}
\end{figure}

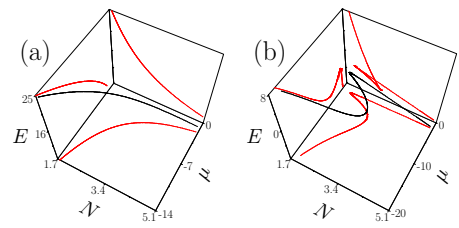
\begin{figure}[tbp]
\caption{(Color online) The soliton family in terms of the
dependence $E=E(\protect\mu ,N)$, and its projections on the
planes $(E,\protect\mu )$, $(E,N)$, and $(\protect\mu ,N)$ for
$p=0$ (a) and $p=3$ (b).} \label{Fig2}
\end{figure}

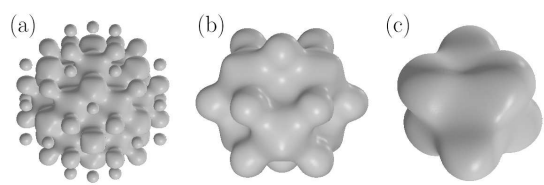
\begin{figure}[tbp]
\caption{Isosurface plots of unstable (a) and stable (b) and (c)
3D solitons. Here $p=3$; (a) $N=2.4$, $A=1.8$, (b) $N=2.04$,
$A=2.2$, and (c) $N=2.4$, $A=3$.} \label{Fig3}
\end{figure}

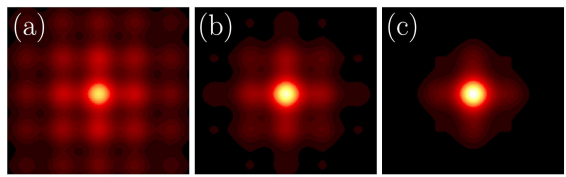
\begin{figure}[tbp]
\caption{(Color online) Integrated through views along the
$z$-axis of the solitons shown in Fig. 3.} \label{Fig4}
\end{figure}

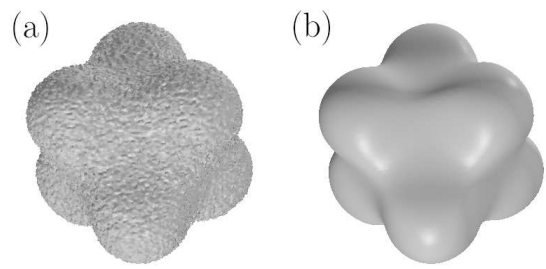
\begin{figure}[tbp]
\caption{ Isosurface plots showing self-cleaning of a stable soliton
corresponding to $p=3$ and $N=2.4$, initially perturbed by white noise. (a)
Input at $t=0$; (b) output at $t=50$.}
\label{Fig5}
\end{figure}

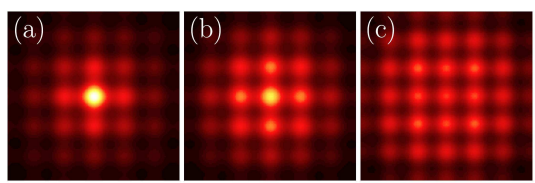
\begin{figure}[tbp]
\caption{(Color online) Integrated through views along the
$z$-axis of an unstable soliton that decays due to a uniform
norm-reducing perturbation with $\protect\epsilon =0.01$, for
$p=3$. (a) $t=0$, $A=1.78$; (b) $t=50$, $A=0.6505$,(c) $t=70$,
$A=0.2075$.} \label{Fig6}
\end{figure}

\end{document}